\documentclass[12pt]{article}
\usepackage{amssymb}

\usepackage{epsfig}
\usepackage{amsmath}

\begin{document}

\title{A stochastic representation for the Poisson-Vlasov equation}
\author{R. Vilela Mendes\thanks{%
CMAF, Complexo Interdisciplinar, Universidade de Lisboa, Av. Gama Pinto, 2 -
1649-003 Lisboa (Portugal), e-mail: vilela@cii.fc.ul.pt;
http://label2.ist.utl.pt/vilela/} \thanks{%
Centro de Fus\~{a}o Nuclear, Instituto Superior T\'{e}cnico, Av. Rovisco
Pais , Lisboa, Portugal} and Fernanda Cipriano\thanks{%
GFM and FCT-Universidade Nova de Lisboa, Complexo Interdisciplinar, Av. Gama
Pinto, 2 - 1649-003 Lisboa (Portugal), e-mail: cipriano@cii.fc.ul.pt}}
\date{}
\maketitle

\begin{abstract}
A stochastic representation for the solutions of the Poisson-Vlasov equation
is obtained. The representation involves both an exponential and a branching
process. The stochastic representation, besides providing an alternative
existence proof and an intuitive characterization of the solutions, may also
be used to obtain an intrinsic definition of the fluctuations.
\end{abstract}

\section{\textbf{Introduction}}

The solutions of linear elliptic and parabolic equations, both with Cauchy
and Dirichlet boundary conditions, have a probabilistic interpretation,
which not only provides intuition on the nature of the problems described by
these equations, but is also quite useful in the proof of general theorems.
This is a very classical field which may be traced back to the work of
Courant, Friedrichs and Lewy \cite{Courant} in the 20's. In spite of the
pioneering work of McKean \cite{McKean}, the question of whether useful
probabilistic representations could also be found for a large class of
nonlinear equations remained an essentially open problem for many years.

It was only in the 90's that, with the work of Dynkin\cite{Dynkin1} \cite
{Dynkin2}, such a theory started to take shape. For nonlinear diffusion
processes, the branching exit Markov systems, that is, processes that
involve diffusion and branching, seem to play the same role as Brownian
motion in the linear equations. However the theory is still limited to some
classes of nonlinearities and there is much room for further mathematical
improvement.

Another field, where considerable recent advances were achieved, was the
probabilistic representation of the Fourier transformed Navier-Stokes
equation, first with the work of LeJan and Sznitman\cite{Jan}, later
followed by extensive developments of the Oregon school\cite{Waymire} \cite
{Bhatta1} \cite{Ossiander}. In all cases the stochastic representation
defines a process for which the mean values of some functionals coincide
with the solution of the deterministic equation.

Stochastic representations, in addition to its intrinsic mathematical
relevance, have several practical implications:

(i) They provide an intuitive characterization of the equation solutions;

(ii) They provide a calculation tool which may replace, for example, the
need for very fine integration grids at high Reynolds numbers;

(iii) By associating a stochastic process to the solutions of the equation,
they provide an intrinsic characterization of the nature of the fluctuations
associated to the physical system. In some cases the stochastic process is
essentially unique, in others there is a class of processes with means
leading to the same solution. The physical significance of this feature is
worth exploring.

A field where stochastic representations have not yet been developed (and
where for the practical applications cited above they might be useful) is
the field of kinetic equations for charged fluids. As a first step towards
this goal, a stochastic representation is here constructed for the solutions of
the Poisson-Vlasov equation.

The comments in the final section point towards future work, in particular
on how a stochastic representation may be used for a characterization of
fluctuations, alternative to existing methods. This is what we call \textit{%
the stochastic principle}.

\section{Stochastic representation and existence}

Consider a Poisson-Vlasov equation in 3+1 space-time dimensions 
\begin{equation}
\frac{\partial f}{\partial t}+\stackrel{\rightarrow }{v}\cdot \nabla _{x}f-%
\frac{e}{m}\nabla _{x}\Phi \cdot \nabla _{v}f=0  \label{2.1}
\end{equation}
with 
\begin{equation}
\Delta _{x}\Phi _{f}=-4\pi \left\{ e\int f\left( \stackrel{\rightarrow }{x},%
\stackrel{\rightarrow }{v},t\right) d^{3}v-e\rho _{B}\left( \stackrel{%
\rightarrow }{x}\right) \right\}  \label{2.2}
\end{equation}
$\rho _{B}\left( \stackrel{\rightarrow }{x}\right) $ being a background
charge density.

Passing to the Fourier transform 
\begin{equation}
F\left( \xi ,t\right) =\frac{1}{\left( 2\pi \right) ^{3}}\int d^{6}\eta
f\left( \eta ,t\right) e^{i\xi \cdot \eta }  \label{2.3}
\end{equation}
with $\eta =\left( \stackrel{\rightarrow }{x},\stackrel{\rightarrow }{v}%
\right) $ and $\xi =\left( \stackrel{\rightarrow }{\xi _{1}},\stackrel{%
\rightarrow }{\xi _{2}}\right) \circeq \left( \xi _{1},\xi _{2}\right) $,
one obtains 
\begin{eqnarray}
0 &=&\frac{\partial F\left( \xi ,t\right) }{\partial t}-\stackrel{%
\rightarrow }{\xi _{1}}\cdot \nabla _{\xi _{2}}F\left( \xi ,t\right) 
\label{2.4} \\
&&+\frac{4\pi e^{2}}{m}\int d^{3}\xi _{1}^{^{\prime }}F\left( \xi _{1}-\xi
_{1}^{^{\prime }},\xi _{2},t\right) \frac{\stackrel{\rightarrow }{\xi _{2}}%
\cdot \stackrel{\rightarrow }{\xi _{1}^{^{\prime }}}}{\left| \xi
_{1}^{^{\prime }}\right| ^{2}}\left\{ F\left( \xi _{1}^{^{\prime
}},0,t\right) -\frac{\stackrel{\symbol{126}}{\rho }_{B}\left( \xi
_{1}^{^{\prime }}\right) }{\left( 2\pi \right) ^{3/2}}\right\}   \nonumber
\end{eqnarray}
$\stackrel{\symbol{126}}{\rho }_{B}\left( \xi _{1}^{^{\prime }}\right) $
being the Fourier transform of $\rho _{B}\left( x\right) $. Changing
variables to 
\begin{equation}
\tau =\gamma \left( \left| \xi _{2}\right| \right) t  \label{2.5}
\end{equation}
where $\gamma \left( \left| \xi _{2}\right| \right) $ is a positive
continuous function satisfying 
\[
\begin{array}{lllll}
\gamma \left( \left| \xi _{2}\right| \right) =1 &  & \textnormal{if} &  & \left|
\xi _{2}\right| <1 \\ 
\gamma \left( \left| \xi _{2}\right| \right) \geq \left| \xi _{2}\right|  & 
& \textnormal{if} &  & \left| \xi _{2}\right| \geq 1
\end{array}
\]
leads to 
\begin{eqnarray}
\frac{\partial F\left( \xi ,\tau \right) }{\partial \tau } &=&\frac{%
\stackrel{\rightarrow }{\xi _{1}}}{\gamma \left( \left| \xi _{2}\right|
\right) }\cdot \nabla _{\xi _{2}}F\left( \xi ,\tau \right) -\frac{4\pi e^{2}%
}{m}\int d^{3}\xi _{1}^{^{\prime }}F\left( \xi _{1}-\xi _{1}^{^{\prime
}},\xi _{2},\tau \right)   \nonumber \\
&&\times \frac{\stackrel{\rightarrow }{\xi _{2}}\cdot \stackrel{\symbol{94}}{%
\xi _{1}^{^{\prime }}}}{\gamma \left( \left| \xi _{2}\right| \right) \left|
\xi _{1}^{^{\prime }}\right| }\left\{ F\left( \xi _{1}^{^{\prime }},0,\tau
\right) -\frac{\stackrel{\symbol{126}}{\rho }_{B}\left( \xi _{1}^{^{\prime
}}\right) }{\left( 2\pi \right) ^{3/2}}\right\}   \label{2.6}
\end{eqnarray}
with $\stackrel{\symbol{94}}{\xi _{1}}=\frac{\stackrel{\rightarrow }{\xi _{1}%
}}{\left| \xi _{1}\right| }$. Eq.(\ref{2.6}) written in integral form, is 
\begin{eqnarray}
F\left( \xi ,\tau \right)  &=&e^{\tau \frac{\stackrel{\rightarrow }{\xi _{1}}%
}{\gamma \left( \left| \xi _{2}\right| \right) }\cdot \nabla _{\xi
_{2}}}F\left( \xi _{1},\xi _{2},0\right) -\frac{4\pi e^{2}}{m}\int_{0}^{\tau
}dse^{\left( \tau -s\right) \frac{\stackrel{\rightarrow }{\xi _{1}}}{\gamma
\left( \left| \xi _{2}\right| \right) }\cdot \nabla _{\xi _{2}}}  \label{2.7}
\\
&&\times \int d^{3}\xi _{1}^{^{\prime }}F\left( \xi _{1}-\xi _{1}^{^{\prime
}},\xi _{2},s\right) \frac{\stackrel{\rightarrow }{\xi _{2}}\cdot \stackrel{%
\symbol{94}}{\xi _{1}^{^{\prime }}}}{\gamma \left( \left| \xi _{2}\right|
\right) \left| \xi _{1}^{^{\prime }}\right| }\left\{ F\left( \xi
_{1}^{^{\prime }},0,s\right) -\frac{\stackrel{\symbol{126}}{\rho }_{B}\left(
\xi _{1}^{^{\prime }}\right) }{\left( 2\pi \right) ^{3/2}}\right\}  
\nonumber
\end{eqnarray}

For convenience, a stochastic representation is going to be written for the
following function 
\begin{equation}
\chi \left( \xi _{1},\xi _{2},\tau \right) =e^{-\lambda \tau }\frac{F\left(
\xi _{1},\xi _{2},\tau \right) }{h\left( \xi _{1}\right) }  \label{2.8}
\end{equation}
with $\lambda $ a constant and $h\left( \xi _{1}\right) $ a positive
function to be specified later on. The integral equation for $\chi \left(
\xi _{1},\xi _{2},\tau \right) $ is 
\begin{eqnarray}
\chi \left( \xi _{1},\xi _{2},\tau \right) &=&e^{-\lambda \tau }\chi \left(
\xi _{1},\xi _{2}+\tau \frac{\xi _{1}}{\gamma \left( \left| \xi _{2}\right|
\right) },0\right) -\frac{8\pi e^{2}}{m\lambda }\frac{\left( \left| \xi
_{1}\right| ^{-1}h*h\right) \left( \xi _{1}\right) }{h\left( \xi _{1}\right) 
}\int_{0}^{\tau }ds\lambda e^{-\lambda s}  \nonumber \\
&&\times \int d^{3}\xi _{1}^{^{\prime }}p\left( \xi _{1},\xi _{1}^{^{\prime
}}\right) \chi \left( \xi _{1}-\xi _{1}^{^{\prime }},\xi _{2}+s\frac{\xi _{1}%
}{\gamma \left( \left| \xi _{2}\right| \right) },\tau -s\right)  \nonumber \\
&&\times \frac{\stackrel{\rightarrow }{\xi _{2}}\cdot \stackrel{\symbol{94}}{%
\xi _{1}^{^{\prime }}}}{\gamma \left( \left| \xi _{2}\right| \right) }%
\left\{ \frac{1}{2}e^{\lambda \left( \tau -s\right) }\chi \left( \xi
_{1}^{^{\prime }},0,\tau -s\right) -\frac{1}{2}\frac{\stackrel{\symbol{126}}{%
\rho }_{B}\left( \xi _{1}^{^{\prime }}\right) }{\left( 2\pi \right)
^{3/2}h\left( \xi _{1}^{^{\prime }}\right) }\right\}  \label{2.9}
\end{eqnarray}
with 
\begin{equation}
\left( \left| \xi _{1}\right| ^{-1}h*h\right) =\int d^{3}\xi _{1}^{^{\prime
}}\left| \xi _{1}^{^{\prime }}\right| ^{-1}h\left( \xi _{1}-\xi
_{1}^{^{\prime }}\right) h\left( \xi _{1}^{^{\prime }}\right)  \label{2.10}
\end{equation}
and 
\begin{equation}
p\left( \xi _{1},\xi _{1}^{^{\prime }}\right) =\frac{\left| \xi
_{1}^{^{\prime }}\right| ^{-1}h\left( \xi _{1}-\xi _{1}^{^{\prime }}\right)
h\left( \xi _{1}^{^{\prime }}\right) }{\left( \left| \xi _{1}\right|
^{-1}h*h\right) }  \label{2.11}
\end{equation}

Eq.(\ref{2.9}) has a stochastic interpretation as an exponential process
(with a time shift in the second variable) plus a branching process. $%
p\left( \xi _{1},\xi _{1}^{^{\prime }}\right) d^{3}\xi _{1}^{^{\prime }}$ is
the probability that, given a $\xi _{1}$ mode, one obtains a $\left( \xi
_{1}-\xi _{1}^{^{\prime }},\xi _{1}^{^{\prime }}\right) $ branching with $%
\xi _{1}^{^{\prime }}$ in the volume $\left( \xi _{1}^{^{\prime }},\xi
_{1}^{^{\prime }}+d^{3}\xi _{1}^{^{\prime }}\right) $. $\chi \left( \xi
_{1},\xi _{2},\tau \right) $ is computed from the expectation value of a
multiplicative functional associated to the processes. Convergence of the
multiplicative functional hinges on the fulfilling of the following
conditions :

(A) $\left| \frac{F\left( \xi _{1},\xi _{2},0\right) }{h\left( \xi
_{1}\right) }\right| \leq 1$

(B) $\left| \frac{\stackrel{\symbol{126}}{\rho }_{B}\left( \xi _{1}\right) }{%
\left( 2\pi \right) ^{3/2}h\left( \xi _{1}\right) }\right| \leq 1$

(C) $\left( \left| \xi _{1}\right| ^{-1}h*h\right) \leq h\left( \xi
_{1}\right) $

Condition (C) is satisfied, for example, for 
\begin{equation}
h\left( \xi _{1}\right) =\frac{c}{\left( 1+\left| \xi _{1}\right|
^{2}\right) ^{2}}\hspace{1cm}\mathnormal{and}\hspace{1cm}c\leq \frac{1}{4\pi 
}  \label{2.12}
\end{equation}
Indeed computing $\frac{1}{h\left( \xi _{1}\right) }\left( \left| \xi
_{1}\right| ^{-1}h*h\right) $ one obtains 
\begin{equation}
\frac{1}{h\left( \xi _{1}\right) }\left( \left| \xi _{1}\right|
^{-1}h*h\right) =4\pi c\int_{0}^{\infty }dr\frac{r}{\left( 1+r^{2}\right)
^{2}}\frac{\left( 1+\left| \xi _{1}\right| ^{2}\right) ^{2}}{\left( 1+\left(
\left| \xi _{1}\right| -r\right) ^{2}\right) \left( 1+\left( \left| \xi
_{1}\right| +r\right) ^{2}\right) }  \label{2.13}
\end{equation}
This integral is bounded by a constant for all $\left| \xi _{1}\right| $,
therefore, choosing $c$ sufficiently small, condition (C) is satisfied.

Once $h\left( \xi _{1}\right) $ consistent with (C) is found, conditions (A)
and (B) only put restrictions on the initial conditions and the background
charge. Now one constructs the stochastic process $X\left( \xi _{1},\xi
_{2},\tau \right) $.

Because $e^{-\lambda \tau }$ is the survival probability during time $\tau $
of an exponential process with parameter $\lambda $ and $\lambda e^{-\lambda
s}ds$ the decay probability in the interval $\left( s,s+ds\right) $, $\chi
\left( \xi _{1},\xi _{2},\tau \right) $ in Eq.(\ref{2.9}) is obtained as the
expectation value of a multiplicative functional for the following
backward-in-time process:

Starting at $\left( \xi _{1},\xi _{2},\tau \right) $, a particle lives for
an exponentially distributed time $s$ up to time $\tau -s$. At its death a
coin $l_{s}$ (probabilities $\frac{1}{2},\frac{1}{2}$) is tossed. If $%
l_{s}=0 $ two new particles are born at time $\tau -s$ with Fourier modes $%
\left( \xi _{1}-\xi _{1}^{^{\prime }},\xi _{2}+s\frac{\xi _{1}}{\gamma
\left( \left| \xi _{2}\right| \right) }\right) $ and $\left( \xi
_{1}^{^{\prime }},0\right) $ with probability density $p\left( \xi _{1},\xi
_{1}^{^{\prime }}\right) $. If $l_{s}=1$ only the $\left( \xi _{1}-\xi
_{1}^{^{\prime }},\xi _{2}+s\frac{\xi _{1}}{\gamma \left( \left| \xi
_{2}\right| \right) }\right) $ particle is born and the process also samples
the background charge at $\stackrel{\symbol{126}}{\rho }_{B}\left( \xi
_{1}^{^{\prime }}\right) $. Each one of the newborn particles continues its
backward-in-time evolution, following the same death and birth laws. When
one of the particles of this tree reaches time zero it samples the initial
condition. The multiplicative functional of the process is the product of
the following contributions:

- At each branching point where two particles are born, the coupling
constant is 
\begin{equation}
g_{2}\left( \xi _{1},\xi _{1}^{^{\prime }},s\right) =-e^{\lambda \left( \tau
-s\right) }\frac{8\pi e^{2}}{m\lambda }\frac{\left( \left| \xi _{1}\right|
^{-1}h*h\right) \left( \xi _{1}\right) }{h\left( \xi _{1}\right) }\frac{%
\stackrel{\rightarrow }{\xi _{2}}\cdot \stackrel{\symbol{94}}{\xi
_{1}^{^{\prime }}}}{\gamma \left( \left| \xi _{2}\right| \right) }
\label{2.14}
\end{equation}

- When only one particle is born and the process samples the background
charge, the coupling is 
\begin{equation}
g_{1}\left( \xi _{1},\xi _{1}^{^{\prime }}\right) =\frac{8\pi e^{2}}{%
m\lambda }\frac{\left( \left| \xi _{1}\right| ^{-1}h*h\right) \left( \xi
_{1}\right) }{h\left( \xi _{1}\right) }\frac{\stackrel{\symbol{126}}{\rho }%
_{B}\left( \xi _{1}^{^{\prime }}\right) }{\left( 2\pi \right) ^{3/2}h\left(
\xi _{1}^{^{\prime }}\right) }\frac{\stackrel{\rightarrow }{\xi _{2}}\cdot 
\stackrel{\symbol{94}}{\xi _{1}^{^{\prime }}}}{\gamma \left( \left| \xi
_{2}\right| \right) }  \label{2.15}
\end{equation}

- When one particle reaches time zero and samples the initial condition the
coupling is 
\begin{equation}
g_{0}\left( \xi _{1},\xi _{2}\right) =\frac{F\left( \xi _{1},\xi
_{2},0\right) }{h\left( \xi _{1}\right) }  \label{2.16}
\end{equation}

The multiplicative functional is the product of all these couplings for each
realization of the process $X\left( \xi _{1},\xi _{2},\tau \right) $, this
process being obtained as the limit of the following iterative process 
\begin{eqnarray*}
X^{\left( k+1\right) }\left( \xi _{1},\xi _{2},\tau \right) &=&\chi \left(
\xi _{1},\xi _{2}+\tau \frac{\xi _{1}}{\gamma \left( \left| \xi _{2}\right|
\right) },0\right) \mathbf{1}_{\left[ s>\tau \right] }+g_{2}\left( \xi
_{1},\xi _{1}^{^{\prime }},s\right) \\
&&\times X^{\left( k\right) }\left( \xi _{1}-\xi _{1}^{^{\prime }},\xi _{2}+s%
\frac{\xi _{1}}{\gamma \left( \left| \xi _{2}\right| \right) },\tau
-s\right) X^{\left( k\right) }\left( \xi _{1}^{^{\prime }},0,\tau -s\right) 
\mathbf{1}_{\left[ s<\tau \right] }\mathbf{1}_{\left[ l_{s}=0\right] } \\
&&+g_{1}\left( \xi _{1},\xi _{1}^{^{\prime }}\right) X^{\left( k\right)
}\left( \xi _{1}^{^{\prime }},0,\tau -s\right) \mathbf{1}_{\left[ s<\tau
\right] }\mathbf{1}_{\left[ l_{s}=1\right] }
\end{eqnarray*}
Then, $\chi \left( \xi _{1},\xi _{2},\tau \right) $ is the expectation value
of the functional. 
\begin{equation}
\chi \left( \xi _{1},\xi _{2},\tau \right) =\Bbb{E}\left\{ \Pi \left(
g_{0}g_{0}^{^{\prime }}\cdots \right) \left( g_{1}g_{1}^{^{\prime }}\cdots
\right) \left( g_{2}g_{2}^{^{\prime }}\cdots \right) \right\}  \label{2.17}
\end{equation}

For example, for the realization in Fig.1 the contribution to the
multiplicative functional is 
\begin{eqnarray*}
&&g_{2}\left( \xi _{1},\xi _{1}^{^{\prime }},\tau -s_{1}\right) g_{2}\left(
\xi _{1}^{^{\prime }},\xi _{1}^{^{\prime \prime \prime }},\tau -s_{3}\right)
g_{1}\left( \xi _{1}-\xi _{1}^{^{\prime }},\xi _{1}^{^{^{\prime \prime
}}}\right) \\
&&\times g_{0}\left( \xi _{1}-\xi _{1}^{^{\prime }}-\xi _{1}^{^{\prime
\prime }},k_{2}\right) g_{0}\left( \xi _{1}^{^{\prime }}-\xi _{1}^{^{\prime
\prime \prime }},k_{3}\right) g_{0}\left( \xi _{1}^{^{^{\prime \prime \prime
}}},0\right)
\end{eqnarray*}

\begin{figure}[tbh]
\begin{center}
\psfig{figure=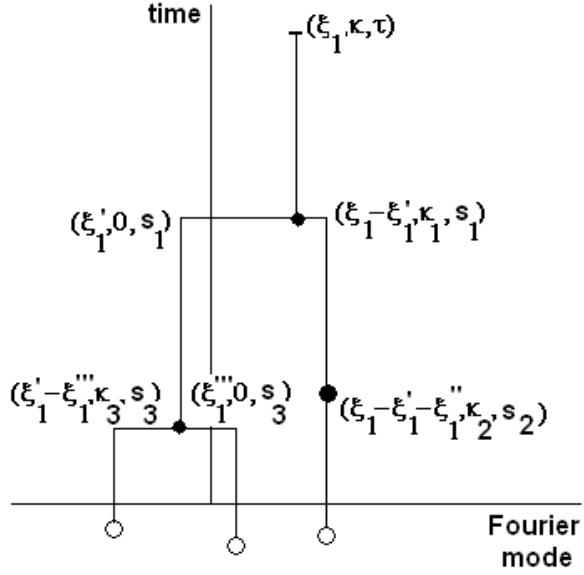,width=11truecm}
\end{center}
\caption{A sample path of the stochastic process}
\end{figure}

and 
\[
\begin{array}{lll}
k & = & \xi _{2} \\ 
k_{1} & = & k+s_{1}\frac{\xi _{1}}{\gamma \left( \left| \xi _{2}\right|
\right) } \\ 
k_{2} & = & k_{1}+\left( s_{2}-s_{1}\right) \frac{\left( \xi _{1}-\xi
_{1}^{^{\prime }}\right) }{\gamma \left( \left| k_{1}\right| \right) } \\ 
k_{3} & = & \left( s_{3}-s_{1}\right) \xi _{1}^{^{\prime }}
\end{array}
\]
With the conditions (A) and (B), choosing 
\[
\lambda =\frac{8\pi e^{2}}{m} 
\]
and 
\[
c\leq e^{-\lambda \tau }\frac{1}{4\pi } 
\]
the absolute value of all coupling constants is bounded by one. The
branching process, being identical to a Galton-Watson process, terminates
with probability one and the number of inputs to the functional is finite
(with probability one). With the bounds on the coupling constants, the
multiplicative functional is bounded by one in absolute value almost surely.

Once a stochastic representation is obtained for $\chi \left( \xi _{1},\xi
_{2},\tau \right) $, one also has, by (\ref{2.8}), a stochastic
representation for the solution of the Fourier-transformed Poisson-Vlasov
equation. The results are summarized in the following :

\textbf{Theorem 2.1}\textit{\ - There is a stochastic representation for the
Fourier-transformed solution of the Poisson-Vlasov equation }$F\left( \xi
_{1},\xi _{2},t\right) $\textit{\ for any arbitrary finite value of the
arguments, provided the initial conditions at time zero and the background
charge satisfy the boundedness conditions (A) and (B).}

As a corollary one also infers an existence result for (arbitrarily large)
finite time. Notice that existence by the stochastic representation method
requires only boundedness conditions on the initial conditions and
background charge and not any strict smoothness properties.

\section{Fluctuations and the stochastic principle. A comment}

In the past, the fluctuation spectrum of charged fluids was studied either
by the BBGKY hierarchy derived from the Liouville or Klimontovich equations,
with some sort of closure approximation, or by direct approximations to the
N-body partition function or by models of dressed test particles, etc. (see
reviews in \cite{Oberman} \cite{Krommes}). Alternatively, by linearizing the
Vlasov equation about a stable solution and diagonalizing the Hamiltonian, a
canonical partition function may be used to compute correlation functions 
\cite{Morrison}.

However, one should remember that, as a model for charged fluids, the Vlasov
equation is just a mean-field collisionless theory. Therefore, it is
unlikely that, by itself, it will contain full information on the
fluctuation spectrum. Kinetic and fluid equations are obtained from the full
particle dynamics in the 6N-dimensional phase-space by a chain of
reductions. Along the way, information on the actual nature of fluctuations
and turbulence may have been lost. An accurate model of turbulence may exist
at some intermediate (mesoscopic) level, but not necessarily in the final
mean-field equation.

When a stochastic representation is constructed, one obtains a process for
which the mean value is the solution of the mean-field equation. The process
itself contains more information. This does not mean, of course, that the
process is an accurate mesoscopic model of Nature, because we might be
climbing up a path different from the one that led us down from the particle
dynamics.

Nevertheless, insofar as the stochastic representation is qualitatively
unique and related to some reasonable iterative process\footnote{%
Representations as those constructed for the Navier-Stokes equation and the
one in this paper may be looked at as a stochastic version of Picard
iteration}, it provides a surrogate mesoscopic model from which fluctuations
are easily computed. This is what we refer to as \textit{the stochastic
principle}. At the minimum, one might say that the stochastic principle
provides another closure procedure.

\end{document}